\documentclass[11pt]{elsarticle}
\usepackage{geometry}
\usepackage{graphicx}

\usepackage{epsfig}
\usepackage{epstopdf}

\usepackage{amsfonts,amsmath,amssymb,amscd}

\usepackage[figuresright]{rotating}

\usepackage{caption}
\usepackage{subcaption}

\usepackage[none]{hyphenat}

\usepackage{lineno}
\usepackage{hyperref} 

\biboptions{round,sort&compress}

\newcommand{\pf}{{\bf Proof : }}
\newtheorem{definition}{Definition}
\newtheorem{theorem}{Theorem}[section]
\newtheorem{lemma}{Lemma}[section]

\newtheorem{example}{Example}[section]

\newcommand{\F}{\mathbb{F}}
\newcommand{\Z}{\mathbb{Z}}

\newcommand{\xn}{x^n - 1}

\begin{document}
\sloppy
\begin{frontmatter}

\title{On cyclic DNA codes over the Ring $\Z_4 + u \Z_4$ }

\author{Sukhamoy Pattanayak and Abhay Kumar Singh$^*$}

\address{Department of Applied Seciences, Indian School of Mines Dhanbad, India\\
	Email: sukhamoy88@gmail.com\\
	$^*$singh.ak.am@ismdhanbad.ac.in}

\begin{abstract}
In this paper, we study the theory for constructing DNA cyclic codes of odd length over $\Z_4[u]/\langle u^2 \rangle$ which play an important role in DNA computing. Cyclic codes of odd length over $\Z_4 + u \Z_4$ satisfy the reverse constraint and the reverse-complement constraint are studied in this paper. The structure and existence of such codes are also studied. The paper concludes with some DNA example obtained via the family of cyclic codes.
\end{abstract}


\end{frontmatter}
Mathematics Subject Classification  94B05. 94B15

\section{Introduction}
%The help is needed \cite{shaw1969}
Deoxyribonucleic acid (DNA) is a nucleic acid containing the genetic instructing used as the carrier of genetic information in all living organisms. DNA is formed by the strands and each strands is sequence consists of four nucleotides; two purines: adenine (A) and guanine (G), and two pyrimidines: thymine (T) and cytosine (C). The two strands of DNA are linked with a rule that are name as Watson-Crick complement (WCC). According to WCC rule; every (A) is linked with a (T), and every (C) with a (G), and vice versa. We write this is as $\overline{A}=T, \overline{T}=A, \overline{G}=C$ and $\overline{C}=G$. For example if $x=(GCATAG)$, then its complement is $\overline{x}=(CGTATC)$. \\
DNA computing links genetic data analysis with scientific computation in order to handle computationally difficult problems. Leonard Adleman [\cite{la}] introduced an experiment involving the use of DNA molecules to solve a hard computational problem in a test tube. His study was based on the WCC property of DNA strands. Several paper have discussed different techniques to construct a set of DNA codewords that are unlikely to form undesirable bonds with each other by hybridization. Four different constraints on DNA codes are considered as follows:
\begin{enumerate}[{\rm (i)}]
	\item The \emph{Hamming constraint}: For any two codewords $x,y \in C, H(x,y)\geq d$ with $x\neq y$, for some minimum distance $d$.
	\item The \emph{reverse constraint}: For any two codewords $x,y \in C, H(x^r,y)\geq d$, where $x^r$ s the reverse of a codeword $x$.
	\item The \emph{reverse-complement constraint}: $ H(x^{rc},y)\geq d$ for all $x,y \in C$.
	\item The \emph{fixed GC-content constraint}: For any codeword $x \in C$ the same number of G and C elements.
\end{enumerate}
The constraints (i) to (iii) is to avoid undesirable hybridization between different strands. The fixed GC-content which ensures all codewords have similar thermodynamic characteristic.\\
Cyclic codes over finite rings played a very important role in the area of error correcting codes [[\cite{ao}], [\cite{yk}], [\cite{absi}], [\cite{vz}], [\cite{bn}], [\cite{d}]]. Since then, the construction of DNA cyclic codes have been discussed by several Authors. Gaborit and King in [\cite{pd}] discussed linear construction of DNA codes. In [\cite{tax}], DNA codes over finite field with four elements were studied by Abualrub et al. Later, Siap et al. studied DNA codes over the finite ring $\F_2[u]/<u^2-1>$ with four element in [\cite{ita}]. In [\cite{jl}], Liang and Wang discussed cyclic DNA codes over four element ring $\F_2+u\F_2$. Yildiz and Siap in [\cite{bi}] studied DNA codes over the ring $\F_2[u]/<u^4-1>$ with 16 elements. \\
Cyclic code over the ring $\Z_4+u\Z_4$ have been discussed in series of papers [[\cite{yk}], [\cite{bm}], [\cite{bn}]]. Here, we study a family of DNA cyclic codes of a finite ring with 16 elements. In this paper, we also study cyclic codes of odd lengths over $\Z_4+u\Z_4$ satisfy the reverse constraint and the reverse-complement constraint.  The sequence of paper is structured as follows: In the second section we discuss the structure of the ring $\Z_4[u]/<u^2>$ and present a description and basic definition of cyclic DNA codes over the ring. Also we establish a 1-1 correspondence $\theta$ between DNA nucleotide pair and the 16 elements of the ring $\Z_4[u]/<u^2>$ and describe cyclic codes of odd length over $\Z_4+u\Z_4$. We study cyclic codes satisfy the reverse constraint over $\Z_4+u\Z_4$ in section 3. In section 4, we also discuss cyclic codes satisfy the reverse-complement constraint over such ring. Moreover, we define the Lee weight related to such codes and give the binary image of the cyclic DNA code in section 5. In section 6, by applying the
theory proved in the previous sections, we present some cyclic DNA codes over the ring $\Z_4[u]/<u^2>$ together with
their images. Section 7 concludes the paper.

\section{Preliminaries}
Let $R$ be the commutative, characteristic 4 ring $\Z_4+u\Z_4=\{a+ub\vert a,b \in \Z_4 \}$ with $u^2=0$. $R$ can also be thought of as the quotient ring $ \Z_4[u]/\langle u^2 \rangle$.
%\begin{center}
%	$1, 3, 1 + u, 1 + 2u, 1 + 3u, 3 + u, 3 + 2u, 3 + 3u$,
%\end{center}
%and the non-units are
%\begin{center}
%	$0, 2, u, 2u, 2+ u, 2 + 2u, 3u, 2 + 3u$.
%\end{center}
%R has six ideals in all listed below:
%\begin{center}
%	$\{0\},\langle u \rangle,\langle 2 \rangle,\langle 2u \rangle,\langle 2+u \rangle,\langle 2,u \rangle$.
%\end{center}
$R$ is a non-principal local ring with $\langle 2,u \rangle$ as its unique maximal ideal. A commutative ring is called a chain ring if its ideals form a chain under the relation of inclusion. Here, $R$ is a non-chain ring. \\
A linear code $C$ of length $n$ over $R$ is a $R$-submodule of $R^n$. An element of $C$ is called a codeword.
A code of length $n$ is cyclic if the code is invariant under the automorphism $\sigma$ which has
\begin{center}
	$\sigma(c_0, c_1, \cdots , c_{n-1}) = (c_{n-1}, c_0, \cdots , c_{n-2}).$
\end{center}
It is well known that a cyclic code of length $n$ over $R$ can be identified with an ideal in the quotient ring $R[x]/\langle x^n-1\rangle$ via the $R$-module isomorphism
as follows:
\begin{center}
	$R^n \longrightarrow R[x]/\langle x^n-1\rangle$\\
	$(c_0,c_1,\cdots,c_{n-1}) \mapsto c_0+c_1x+\cdots+c_{n-1}x^{n-1} (\text{mod}\langle x^n-1\rangle) $
\end{center}
DNA occurs in sequences, represented by sequences of nucleotides $\{A,T,G,C\}$. We define a DNA code of length $n$ to be a set of codewords $(x_0,x_1,\cdots, x_{n-1})$ where $x_i \in \{A,T,G,C\}$. These codewords must satisfy the four constraints mentioned above introduction. In this paper, we have 16 pairs constructed by four basic nucleotides $A,T,G ~\text{and}~C$ such as 
\begin{center}
	$AA,TT,GG,CC,AT,TA,GC,CG,GT,TG,AC,CA,CT,TC,AG,GA$.
\end{center}
Since the ring $R$ is of the cardinality 16, then we define the map $\theta$ which gives a one-to-one correspondence between the elements of $R$ and the 16 codons over the alphabet $\{A,T,G,C\}^2$, which is given in Table 1. The codons satisfy the Watson-Crick complement.
\begin{center}
	{\bf Table 1.} Identifying Codons with the Elements of the Ring $R$.\\~\\
	\begin{tabular}{ l  c  c  c  c  c  c  c  c }
		\hline
		
		$AA$ & $0$ & $TT$ & $1+u$ & $GG$ & $1$ & $CC$ & $u$ \\
		$AT$ & $2$ & $TA$ & $3+u$ & $GC$ & $3$ & $CG$ & $2+u$ \\
		$GT$ & $2u$ & $CA$ & $1+3u$ & $AC$ & $3u$ & $TG$ & $1+2u$ \\
		$CT$ & $2+3u$ & $GA$ & $3+2u$ & $AG$ & $2+2u$ & $TC$ & $3+3u$  \\
		\hline
		
	\end{tabular}
\end{center}
Let $x=x_0x_1 \cdots x_{n-1} \in R^n$ be a vector. The reverse of x is defined as $x^r = x_{n-1}x_{n-2} \cdots x_1x_0$, the complement of x
is $x^c =\overline{x_0}~\overline{x_1}\cdots \overline{x_{n-1}} $, and the reverse-complement, also called the Watson-Crick complement (WCC) is defined as $x^{rc} =\overline{x_{n-1}}~\overline{x_{n-2}} \cdots \overline{x_1}~\overline{x_0} $.
\begin{definition}
	A linear code $C$ of length $n$ over $R$ is said to be \bfseries{reversible} if $x^r \in C~~ \forall~ x \in C$, \bfseries{complement} if $x^c \in C~~ \forall~ x \in C$ and \bfseries{reversible-complement} if $x^{rc} \in C~~ \forall~ x \in C$.	
\end{definition}
\begin{definition}
	A cyclic code $C$ of length $n$ is called DNA code over$R$ if
	\begin{enumerate}[{\rm (1)}]
		\item $C$ is cyclic code, i.e. $C$ is an ideal of $R[x]/\langle x^n-1\rangle$;
		\item For any codeword $x \in C, x \neq x^{rc}$	and $x^{rc} \in C$.
	\end{enumerate}
\end{definition}
For each polynomial $f(x)=a_0+a_1x+\cdots+a_rx^r$ with $a_r\neq 0$, we define the reciprocal of $f(x)$ to be the polynomial
\begin{center}
	$f^*(x)=x^rf(1/x)=a_rx^r+a_{r-1}x+\cdots+a_0x^r$.
\end{center}
It is easy to see that $~deg~(f ^*(x)) \leq deg~(f (x))$ and if $a_0 \neq 0$, then $~deg~(f ^*(x)) = deg~(f (x))$. $f(x)$ is called a self-reciprocal polynomial if there is a constant $m$ such that $f^*(x) = mf(x)$.\\
The structure of cyclic code of odd length $n$ over $R$ has been studied, which is
\begin{theorem}
	Let $n$ be an odd integer and $C$ be a cyclic code of length $n$ over $R$. Then
	\begin{enumerate}[{\rm (1)}]
		\item	$C=\langle f_1(x)+2f_2(x)+2uf_{14}(x),uf_3(x)+2uf_4(x) \rangle$ ,
		where $f_2(x)\vert f_1(x)\vert (\xn)$ and $f_4(x)\vert f_3(x)\vert (\xn)$ in $\frac{R[x]}{\langle x^n-1\rangle}$. 
		\item If $f_1(x)=f_4(x)$, then $C=\langle f_1(x)+2f_2(x)+2uf_{14}(x)\rangle$ where $f_2(x)\vert f_1(x)\vert (\xn)$.
	\end{enumerate}	
\end{theorem}

\section{Reversible Codes over $R$}
In this section, we study the reverse constraint on cyclic codes of odd length over $R$. First, we give some useful lemma's which use the following theorems.
\begin{lemma} \cite{absi}
	Let $f(x),g(x),h(x)$ be any three polynomials in $R$ with $~deg~f(x) \geq ~deg~g(x) \geq ~deg~h(x)$. Then
	\begin{enumerate}[{\rm (1)}]
		\item $[f(x)g(x)h(x)]^*=f^*(x)g^*(x)h^*(x)$;
		\item $[f(x)+g(x)+h(x)]^*=f^*(x)+x^{~deg~f-deg~g}g^*(x)+x^{~deg~f-deg~h}h^*(x)$.
	\end{enumerate}
\end{lemma}
\begin{lemma}
	Let $C = \langle f (x)\rangle$ be a cyclic code over $\Z_4$, then $C$ is reversible if and
	only if $f(x)$ is self-reciprocal.
\end{lemma}
\begin{theorem}
	Let $C=\langle f_1(x)+2f_2(x)+2uf_{14}(x)\rangle$ with $f_2(x)\vert f_1(x)\vert \xn$ be a cyclic code of odd length $n$ over $R$. Then $C$ is reversible if and only if
	\begin{enumerate}[{\rm (a)}]
		\item $f_1(x)$ is self-reciprocal;
		\item \begin{enumerate}[{\rm (i)}]
			\item $x^if_2^*(x)=f_2(x)$ and
			\item $x^jf_{14}^*(x)=f_{14}(x)$ or $f_2(x)\vert (2x^jf_{14}^*(x)+2f_{14}(x))$, \\
			where $i=~deg~f_1(x)-deg~f_2(x), j=~deg~f_1(x)-deg~f_{14}(x)$.
		\end{enumerate}
	\end{enumerate}
\end{theorem}
\pf Suppose $C=\langle f_1(x)+2f_2(x)+2uf_{14}(x)\rangle$ is reversible over $R$, then $C$ is reversible over $\Z_4$, then from Lemma 3.2, $f_1(x)$ is self-reciprocal. That implies 
\begin{align}
	(f_1(x)+2f_2(x)+2uf_{14}(x))^* & = f_1^*(x)+2x^if_2^*(x)+2ux^jf_{14}^*(x) \nonumber \\ 
	& = f_1(x)+2x^if_2^*(x)+2ux^jf_{14}^*(x)\\
	& = (f_1(x)+2f_2(x)+2uf_{14}(x))a(x)
\end{align}
where $i=~deg~f_1(x)-deg~f_2(x), j=~deg~f_1(x)-deg~f_{14}(x)$. Comparing the degree of (1) and (2), we get $a(x)=c$, where $c\in R$. Then
\begin{equation}
	f_1(x)+2x^if_2^*(x)+2ux^jf_{14}^*(x)=c.f_1(x)+c.2f_2(x)+c.2uf_{14}(x).
\end{equation}
Multiplying $2u$ both side of (3) we get, $2uf_1(x)=c.2uf_1(x)$. That implies $c=1,1+u,1+2u ~\text{or}~1+3u$. If $c=1$, then by (3) we write 
\begin{equation}
	2x^if_2^*(x)+2ux^jf_{14}^*(x)=2f_2(x)+2uf_{14}(x).
\end{equation}
Multiplying $u$ both side of (4) we get, $2ux^if_2^*(x)=2uf_2(x)$ as $u^2=0$, that implies that $x^if_2^*(x)=f_2(x)$ where $i=~deg~f_1(x)-deg~f_2(x)$. Again using that result on (4) we get, $x^jf_{14}^*(x)=f_{14}(x)$ where $j=~deg~f_1(x)-deg~f_{14}(x)$.\\
For $c=1+u$, then by (3) we have, 
\begin{equation}
	f_1(x)+2x^if_2^*(x)+2ux^jf_{14}^*(x)=f_1(x)+uf_1(x)+2f_2(x)+2uf_2(x)+2uf_{14}(x).
\end{equation} 
Multiplying $u$ both side of (5) we get, $x^if_2^*(x)=f_2(x)$. Again using that result on (5) we write, $2ux^jf_{14}^*(x)+2uf_{14}(x)=uf_1(x)+2uf_2(x)$. Since $f_2(x)\vert f_1(x)$, then we have $(2x^jf_{14}^*(x)+2f_{14}(x)) \in (f_2(x))$. Therefore, $f_2(x)\vert (2x^jf_{14}^*(x)+2f_{14}(x))$.\\
For $c=1+2u$,then by (3) we have, 
\begin{equation}
	f_1(x)+2x^if_2^*(x)+2ux^jf_{14}^*(x)=f_1(x)+2uf_1(x)+2f_2(x)+2uf_{14}(x).
\end{equation} 
Similarly multiplying $u$ both side of (6) we get, $x^if_2^*(x)=f_2(x)$. Using the result on (6) we get, $2x^jf_{14}^*(x)+2f_{14}(x)=2f_1(x)$. Therefore, $f_2(x)\vert (2x^jf_{14}^*(x)+2f_{14}(x))$ as $f_2(x)\vert f_1(x)$. Similarly for $c=1+3u$ we get the same result. \\
On the other hand, we have
\begin{align}
	(f_1(x)+2f_2(x)+2uf_{14}(x))^* & = f_1^*(x)+2x^if_2^*(x)+2ux^jf_{14}^*(x) \nonumber \\ 
	& = f_1(x)+2x^if_2^*(x)+2ux^jf_{14}^*(x)\nonumber \\
	& = (f_1(x)+2f_2(x)+2uf_{14}(x))c \in C, \nonumber
\end{align}
where $c=1,1+u,1+2u ~\text{or}~1+3u$, $i=deg~f_1(x)-deg~f_2(x)$ and $j=deg~f_1(x)-deg~f_{14}(x)$. Hence, $C$ is reversible.
\begin{theorem}
	Let $C=\langle f_1(x)+2f_2(x)+2uf_{14}(x),uf_3(x)+2uf_4(x) \rangle$
	with $f_2(x)\vert f_1(x)\vert \xn$ and $f_4(x)\vert f_3(x)\vert \xn$ in $\frac{R[x]}{\langle x^n-1\rangle}$ be a cyclic code of odd length $n$ over $R$. Then $C$ is reversible if and only if
	\begin{enumerate}[{\rm (a)}]
		\item $f_1(x)$ and $f_3(x)$ are self-reciprocal;
		\item \begin{enumerate}[{\rm (i)}]
			\item $x^if_2^*(x)=f_2(x)$ and
			\item $f_4(x)\vert (2x^jf_{14}^*(x)+2f_{14}(x))$ or $f_4(x)\vert (2x^jf_{14}^*(x)+2f_{14}(x)+2f_2(x))$, \\
			where $i=~deg~f_1(x)-deg~f_2(x), j=~deg~f_1(x)-deg~f_{14}(x)$.
		\end{enumerate}
	\end{enumerate}
\end{theorem}
\pf Suppose $C=\langle f_1(x)+2f_2(x)+2uf_{14}(x),uf_3(x)+2uf_4(x) \rangle$ is reversible. Then $\langle f_1(x)\rangle$ and  $\langle f_3(x)\rangle$ are reversible over $\Z_4$, from Lemma 3.2 $\langle f_1(x)\rangle$ and  $\langle f_3(x)\rangle$ are self-reciprocal. Since $C$ is reversible this implies
\begin{align}
	(f_1(x)+2f_2(x)+2uf_{14}(x))^* & = f_1^*(x)+2x^if_2^*(x)+2ux^jf_{14}^*(x) \nonumber \\ 
	& = f_1(x)+2x^if_2^*(x)+2ux^jf_{14}^*(x)\\
	& = (f_1(x)+2f_2(x)+2uf_{14}(x))a(x) \nonumber \\ & +u (f_3(x)+2 f_4(x))b_1(x) \nonumber \\
	& = (f_1(x)+2f_2(x)+2uf_{14}(x))a(x)\nonumber \\ & +u f_4(x)b(x),
\end{align}
since $f_4(x)\vert f_3(x)$. Here $i=~deg~f_1(x)-deg~f_2(x), j=~deg~f_1(x)-deg~f_{14}(x)$. Comparing the degree of (7) and (8), we have $a(x)=c$, where $c\in R$. Then
\begin{equation}
	f_1(x)+2x^if_2^*(x)+2ux^jf_{14}^*(x)=c.f_1(x)+c.2f_2(x)+c.2uf_{14}(x)+u (f_3(x)+2 f_4(x))b(x)
\end{equation}
Multiplying $2u$ both side of (9) we write, $2uf_1(x)=c.2uf_1(x)$. That implies $c=1,1+u,1+2u ~\text{or}~1+3u$. If $c=1$, then by (9) we have 
\begin{equation}
	f_1(x)+2x^if_2^*(x)+2ux^jf_{14}^*(x)=f_1(x)+2f_2(x)+2uf_{14}(x)+u (f_3(x)+2 f_4(x))b(x).
\end{equation}
Multiplying $u$ both side of (10) we get, $2ux^if_2^*(x)=2uf_2(x)$ as $u^2=0$, that implies that $x^if_2^*(x)=f_2(x)$ where $i=~deg~f_1(x)-deg~f_2(x)$. Again using that result on (10) we write, $2ux^jf_{14}^*(x)+2uf_{14}(x)=u (f_3(x)+2 f_4(x))b(x)$. Since $f_4(x)\vert f_3(x)$, then we have $(2x^jf_{14}^*(x)+2f_{14}(x)) \in (f_4(x))$. Therefore,\\ $f_4(x)\vert (2x^jf_{14}^*(x)+2f_{14}(x))$.\\
For $c=1+u$, then by (9) we get the equation, 
\begin{align}
	f_1(x)+2x^if_2^*(x)+2ux^jf_{14}^*(x) &=f_1(x)+uf_1(x)+2f_2(x)+2uf_2(x)+2uf_{14}(x) \nonumber \\ & +u (f_3(x)+2 f_4(x))b(x).
\end{align} 
Multiplying $u$ both side of (11) we get, $x^if_2^*(x)=f_2(x)$. Again using that result on (11) we write, $2ux^jf_{14}^*(x)+2uf_{14}(x)+2uf_2(x)=uf_1(x)+u (f_3(x)+2 f_4(x))b(x)$. Since $f_4(x)\vert f_3(x)\vert f_1(x)$, then we write $(2x^jf_{14}^*(x)+2f_{14}(x)+2f_2(x)) \in (f_4(x))$. Therefore, $f_4(x)\vert (2x^jf_{14}^*(x)+2f_{14}(x)+2f_2(x))$.\\
Similarly for $c=1+2u$ and $c=1+3u$ we get the same result as $c=1$ and $c=1+u$ respectively. \\
Conversely, for $C$ to be reversible it is sufficient to show that both $(f_1(x)+2f_2(x)+2uf_{14}(x))^*$ and $(uf_3(x)+u2f_4(x))^*$ are in $C$. Since $f_3(x)$ is self-reciprocal and $f_4(x)\vert f_3(x)$ then $uf_4(x)\in C$. Also
\begin{align}
	(f_1(x)+2f_2(x)+2uf_{14}(x))^* & = f_1^*(x)+2x^if_2^*(x)+2ux^jf_{14}^*(x) \nonumber \\ 
	& = (f_1(x)+2f_2(x)+2uf_{14}(x))+2ux^jf_{14}^*(x)+2uf_{14}(x) \nonumber \\ 
	&=(f_1(x)+2f_2(x)+2uf_{14}(x))+uf_4(x)b(x) \in C \nonumber
\end{align}
Therefore, $C$ is reversible.

\section{Cyclic Reversible Complement Codes over $R$}
In this section, cyclic codes of odd lengths over $R$ satisfy the reverse-complement are examined.
First, we give some useful lemmas which can be easily check.
\begin{lemma}
	For any $a \in R$ we have $a+\overline{a}=1+u$.
\end{lemma}
\begin{lemma}
	For any $a,b,c \in R$, then
	\begin{enumerate}[{\rm (i)}]
		\item $\overline{a+b}=\overline{a}+\overline{b}+3(1+u)$;
		\item $\overline{a+b+c}=\overline{a}+\overline{b}+\overline{c}+2(1+u)$.
	\end{enumerate}
\end{lemma}
\begin{lemma}
	For any $a \in \Z_4$ we have $\overline{2ua}+3(1+u)=2au$.
\end{lemma}
\begin{lemma}
	For any $a \in R$, then we have
	\begin{enumerate}[{\rm (i)}]
		\item $\overline{2a}+3(1+u)=2a$;
		\item $\overline{a}+3(1+u)=3a$.
	\end{enumerate}
\end{lemma}

\begin{theorem}
	Let $C=\langle f_1(x)+2f_2(x)+2uf_{14}(x)\rangle$ with $f_2(x)\vert f_1(x)\vert \xn$ be a cyclic code of odd length $n$ over $R$. Then $C$ is a reverse-complement if and only if
	\begin{enumerate}[{\rm (a)}]
		\item $f_1(x)$ is self-reciprocal and $3(1+u)((1-x^n)/(1-x)) \in C$;
		\item \begin{enumerate}[{\rm (i)}]
			\item $x^if_2^*(x)=f_2(x)$ and
			\item $x^jf_{14}^*(x)=f_{14}(x)$ or $f_2(x)\vert (2x^jf_{14}^*(x)+2f_{14}(x))$, \\
			where $i=~deg~f_1(x)-deg~f_2(x), j=~deg~f_1(x)-deg~f_{14}(x)$.
		\end{enumerate}
	\end{enumerate}	
\end{theorem}
\pf Let $C$ be the cyclic code given as in the theorem. Since the zero codeword must be in $C$, by hypothesis, the WCC of it should also be in $C$. But by Lemma 4.1, we have
\begin{equation}
	3~\overline{(0,0,\cdots,0)}=3(1+u,1+u,\cdots,1+u)=3(1+u)\frac{1-x^n}{1-x}\in C.\nonumber
\end{equation}
Now, let
\begin{align}
	&
	f_1(x)=1+g_1^{\prime}x+\cdots+g_{s-1}^{\prime}x^{s-1}+x^s, \nonumber \\
	&
	f_2(x)=1+g_1^{\prime\prime}x+\cdots+g_{r-1}^{\prime\prime}x^{r-1}+x^r, \nonumber \\
	& ~\text{and}~~~ f_{14}(x)=1+g_1^{\prime\prime\prime}x+\cdots+g_{t-1}^{\prime\prime\prime}x^{t-1}+x^t \nonumber,
\end{align}
where $s>r>t$. Then
\begin{multline}
	\lefteqn {f_1(x)+2f_2(x)+2uf_{14}(x)} \\ =  (1+g_1^{\prime}x + \cdots+g_{s-1}^{\prime}x^{s-1}+x^s) +  (2+2g_1^{\prime\prime}x+\cdots+2g_{r-1}^{\prime\prime}x^{r-1}+2x^r) \\ + (2u+2ug_1^{\prime\prime\prime}x+\cdots+2ug_{t-1}^{\prime\prime\prime}x^{t-1}+2ux^t)  \\  = (3+2u)+(g_1^{\prime}+2g_1^{\prime\prime}+2ug_1^{\prime\prime\prime})x + \cdots +  (g_{t-1}^{\prime}+2g_{t-1}^{\prime\prime}+2ug_{t-1}^{\prime\prime\prime})x^{t-1}+ \\ \cdots + (g_t^{\prime}+2g_t^{\prime\prime}+2u^)x^t + (g_{t+1}^{\prime}+2g_{t+1}^{\prime\prime})x^{t+1} + \cdots + (g_{r-1}^{\prime}+2g_{r-1}^{\prime\prime})x^{r-1} \\ + (g_r^{\prime}+2)x^r + g_{r+1}^{\prime}x^{r+1} + \cdots + g_{s-1}^{\prime}x^{s-1} + x^s. \nonumber
\end{multline}
Hence,
\begin{multline}
	\lefteqn (~{f_1(x)+2f_2(x)+2uf_{14}(x)})^{rc} \\ = (1+u)(1+x+\cdots+x^{n-s-2}) + ux^{n-s-1} + \overline{g_{s-1}^{\prime}}x^{n-s} + \cdots + \overline{g_{r+1}^{\prime}}x^{n-r-2} \\ + \overline{(g_r^{\prime}+2)}x^{n-r-1} + \overline{(g_{r-1}^{\prime}+2g_{r-1}^{\prime\prime})}x^{n-r} + \cdots + \overline{(g_{t+1}^{\prime}+2g_{t+1}^{\prime\prime})}x^{n-t-2} + \\ \overline{(g_t^{\prime}+2g_t^{\prime\prime}+2u^)}x^{n-t-1} + \overline{(g_{t-1}^{\prime}+2g_{t-1}^{\prime\prime}+2ug_{t-1}^{\prime\prime\prime})}x^{n-t} + \cdots + \\ \overline{(g_1^{\prime}+2g_1^{\prime\prime}+2ug_1^{\prime\prime\prime})}x^{n-2} + \overline{(3+2u)}x^{n-1} \\ = 
	(1+u)(1+x+\cdots+x^{n-s-2}) + ux^{n-s-1} + \overline{g_{s-1}^{\prime}}x^{n-s} + \cdots + \overline{g_{r+1}^{\prime}}x^{n-r-2} \\ + \overline{g_r^{\prime}}x^{n-r-1} + 2x^{n-r-1} + \overline{g_{r-1}^{\prime}}x^{n-r} + 2g_{r-1}^{\prime\prime}x^{n-r} + \cdots + \overline{g_{t+1}^{\prime}}x^{n-t-2} + \\ 2g_{t+1}^{\prime\prime}x^{n-t-2} + \overline{g_t^{\prime}}x^{n-t-1} + c + 2ux^{n-t-1} + \overline{g_{t-1}^{\prime}}x^{n-t} +  2g_{t-1}^{\prime\prime}x^{n-t} \\ +  2ug_{t-1}^{\prime\prime\prime}x^{n-t} + \cdots + \overline{g_1^{\prime}}x^{n-2} + 2g_1^{\prime\prime}x^{n-2} + 2ug_1^{\prime\prime\prime}x^{n-2} + (2+3u)x^{n-1} \\ =
	(1+u)(1+x+\cdots+x^{n-s-2}) + ux^{n-s-1} + \overline{g_{s-1}^{\prime}}x^{n-s} + \cdots + \overline{g_{r+1}^{\prime}}x^{n-r-2} \\ + \overline{g_r^{\prime}}x^{n-r-1} + \overline{g_{r-1}^{\prime}}x^{n-r} + \overline{g_{t+1}^{\prime}}x^{n-t-2} + \overline{g_t^{\prime}}x^{n-t-1} + \overline{g_{t-1}^{\prime}}x^{n-t} + \cdots +  \overline{g_1^{\prime}}x^{n-2} \\ +  ux^{n-1} + 2x^{n-r-1} + 2g_{r-1}^{\prime\prime}x^{n-r} + \cdots + 2g_{t+1}^{\prime\prime}x^{n-t-2} + 2g_t^{\prime\prime}x^{n-t-1} + 2g_{t-1}^{\prime\prime}x^{n-t} \\ + \cdots +  2g_1^{\prime\prime}x^{n-2} + 2x^{n-1} + 2ux^{n-t-1} + 2ug_{t-1}^{\prime\prime\prime}x^{n-t} + \cdots + 2ug_1^{\prime\prime\prime}x^{n-2} + 2ux^{n-1} \\ =
	(1+u)(1+x+\cdots+x^{n-s-2}) + ux^{n-s-1} + \overline{g_{s-1}^{\prime}}x^{n-s} + \cdots + \overline{g_1^{\prime}}x^{n-2} + ux^{n-1} \\ + 3.2x^{n-r-1} + 3.2g_{r-1}^{\prime\prime}x^{n-r} + \cdots + 3.2g_1^{\prime\prime}x^{n-2} + 3.2x^{n-1} \\ + 3.2ux^{n-t-1} + 3.2ug_{t-1}^{\prime\prime\prime}x^{n-t} +  \cdots + 3.2ug_1^{\prime\prime\prime}x^{n-2} + 3.2ux^{n-1}. \nonumber
\end{multline}
Since $C$ is linear code, we must have
\begin{equation}
	({f_1(x)+2f_2(x)+2uf_{14}(x)})^{rc} + 3(1+u)\frac{1-x^n}{1-x}\in C.\nonumber
\end{equation}
That implies that
\begin{multline}
	({f_1(x)+2f_2(x)+2uf_{14}(x)})^{rc} + 3(1+u)((1-x^n)/(1-x)) \\ = 
	3x^{n-s-1} + (\overline{g_{s-1}^{\prime}}+3(1+u))x^{n-s} + \cdots + (\overline{g_1^{\prime}}+3(1+u))x^{n-2} + 3x^{n-1} \\ + 3.2x^{n-r-1}(1+g_{r-1}^{\prime\prime}x+\cdots+g_1^{\prime\prime}x^{r-1}+x^r) \\ +  3.2ux^{n-t-1}(1+g_{t-1}^{\prime\prime\prime}x+\cdots+g_1^{\prime\prime\prime}x^{t-1}+x^t) \\ =
	3x^{n-s-1} + 3g_{s-1}^{\prime}x^{n-s} + \cdots + 3g_1^{\prime}x^{n-2} + 3x^{n-1} \\ + 3.2x^{n-r-1}(1+g_{r-1}^{\prime\prime}x+\cdots+g_1^{\prime\prime}x^{r-1}+x^r) \\ +  3.2ux^{n-t-1}(1+g_{t-1}^{\prime\prime\prime}x+\cdots+g_1^{\prime\prime\prime}x^{t-1}+x^t) \\ =
	3x^{n-s-1}(1+g_{s-1}^{\prime}x+\cdots+g_1^{\prime}x^{s-1}+x^s)\\ + 3.2x^{n-r-1}(1+g_{r-1}^{\prime\prime}x+\cdots+g_1^{\prime\prime}x^{r-1}+x^r) \\ +  3.2ux^{n-t-1}(1+g_{t-1}^{\prime\prime\prime}x+\cdots+g_1^{\prime\prime\prime}x^{t-1}+x^t) \\ =
	3x^{n-s-1}f_1^*(x) + 3.2x^{n-r-1}f_2^*(x) + 3.2ux^{n-t-1}f_{14}^*(x) \\=
	3x^{n-s-1}(f_1^*(x)+2x^{s-r}f_2^*(x)+2ux^{s-t}f_{14}^*(x)) \nonumber.
\end{multline}
Whence, 
\begin{equation}
	f_1^*(x)+2x^{s-r}f_2^*(x)+2ux^{s-t}f_{14}^*(x) \in C. \nonumber
\end{equation}
So we have,
\begin{equation}
	f_1^*(x)+2x^{s-r}f_2^*(x)+2ux^{s-t}f_{14}^*(x)=(f_1(x)+2f_2(x)+2uf_{14}(x))a(x) \nonumber
\end{equation}
It is easy to see that $a(x)=1,1+u,1+2u ~\text{or}~ 1+3u$. So by the previous Theorem 3.1, $f_1(x)=f_1^*(x)$ i.e, $f_1(x)$ s self-reciprocal. Also we have, $x^if_2^*(x)=f_2(x)$ and $x^jf_{14}^*(x)=f_{14}(x)$ or $f_2(x)\vert (2x^jf_{14}^*(x)+2f_{14}(x))$, \\
where $i=s-r, j=s-t$.\\
On the other hand, let $c(x)\in C$, then $c(x)=(f_1(x)+2f_2(x)+2uf_{14}(x))a(x)$. Since $f_1(x)$ self-reciprocal and also $x^if_2^*(x)=f_2(x)$ and $x^jf_{14}^*(x)=f_{14}(x)$ or $f_2(x)\vert (2x^jf_{14}^*(x)+2f_{14}(x))$, where $i=~deg~f_1(x)-deg~f_2(x)~\text{and}~ j=~deg~f_1(x)-deg~f_{14}(x)$, then we write, 
\begin{align}
	c^*(x) & =((f_1(x)+2f_2(x)+2uf_{14}(x))a(x))^* \nonumber \\ &=(f_1^*(x)+2x^if_2^*(x)+2ux^jf_{14}^*(x))a^*(x) \nonumber \\ 
	& = (f_1(x)+2f_2(x)+2uf_{14}(x))a^*(x) \nonumber \\ 
	& = (f_1(x)+2f_2(x)+2uf_{14}(x)).c.a^*(x), \nonumber
\end{align}
Where $c=1,1+u,1+2u ~\text{or}~ 1+3u$. Therefore $c^*(x) \in C$. \\ 
Since $3(1+u)((1-x^n)/(1-x)) \in C$, we have
\begin{equation}
	3(1+u)\frac{1-x^n}{1-x}=3(1+u)+3(1+u)x+\cdots+3(1+u)x^{n-1} \in C. \nonumber
\end{equation}
Let $c(x)=c_0+c_1x+\cdots+c_mx^m \in C$. As $C$ is a cyclic code of length $n$, we have
\begin{equation}
	x^{n-m-1}c(x)=c_0x^{n-m-1}+c_1x^{n-m}+\cdots+c_mx^{n-1} \in C. \nonumber
\end{equation}
Whence, 
\begin{multline}
	\lefteqn ~ 3(1+u)+3(1+u)x+\cdots+3(1+u)x^{n-m-2}+(c_0+3(1+u))x^{n-m-1} \\ +(c_1+3(1+u))x^{n-m}+\cdots+(c_m+3(1+u))x^{n-1} \\ =
	3(1+u)+3(1+u)x+\cdots+3(1+u)x^{n-m-2}+ 3\overline{c_0}x^{n-m-1} \\ +3\overline{c_1}x^{n-m}+\cdots+ 3\overline{c_m}x^{n-1}  \\ =
	3((1+u)+(1+u)x+\cdots+(1+u)x^{n-m-2}+ \overline{c_0}x^{n-m-1} \\ +\overline{c_1}x^{n-m}+\cdots+ \overline{c_m}x^{n-1}) \in C. \nonumber
\end{multline}
That implies $c^*(x)^{rc} \in C$, therefore, $(c^*(x)^{rc})^*=c(x)^{rc} \in C$. Hence proved.
\begin{theorem}
	Let $C=\langle f_1(x)+2f_2(x)+2uf_{14}(x),uf_3(x)+2uf_4(x) \rangle$ be a cyclic code of odd length $n$ over $R$ with $f_2(x)\vert f_1(x)\vert \xn$ and $f_4(x)\vert f_3(x)\vert \xn$ in $\frac{R[x]}{\langle x^n-1\rangle}$. Then $C$ is reverse-complement if and only if
	\begin{enumerate}[{\rm (a)}]
		\item $3(1+u)((1-x^n)/(1-x)) \in C$ and $f_1(x)$ and $f_3(x)$ are self-reciprocal;
		\item \begin{enumerate}[{\rm (i)}]
			\item $x^if_2^*(x)=f_2(x)$ and
			\item $f_4(x)\vert (2x^jf_{14}^*(x)+2f_{14}(x))$ or $f_4(x)\vert (2x^jf_{14}^*(x)+2f_{14}(x)+2f_2(x))$, \\
			where $i=~deg~f_1(x)-deg~f_2(x), j=~deg~f_1(x)-deg~f_{14}(x)$.
		\end{enumerate}
	\end{enumerate}
\end{theorem}
\pf Suppose Let $C=\langle f_1(x)+2f_2(x)+2uf_{14}(x),uf_3(x)+2uf_4(x) \rangle$ with $f_2(x)\vert f_1(x)\vert \xn$ and $f_4(x)\vert f_3(x)\vert \xn$ in $\frac{R[x]}{\langle x^n-1\rangle}$. Since the zero codeword must be in $C$, by hypothesis, the WCC of it should also be in $C$, i.e,
\begin{equation}
	3~\overline{(0,0,\cdots,0)}=3(1+u,1+u,\cdots,1+u)=3(1+u)\frac{1-x^n}{1-x}\in C.\nonumber
\end{equation}
Proceeding in the same way as previous theorem, we get
\begin{equation}
	f_1^*(x)+2x^{s-r}f_2^*(x)+2ux^{s-t}f_{14}^*(x) \in C. \nonumber
\end{equation}
Where $deg~f_1(x)=s, deg~f_1(x)=r, deg~f_1(x)=t$. 
So we have,
\begin{align}
	f_1^*(x)+2x^{s-r}f_2^*(x)+2ux^{s-t}f_{14}^*(x) & =(f_1(x)+2f_2(x)+2uf_{14}(x))a(x) \nonumber \\
	& +u (f_3(x)+2 f_4(x))b(x)
\end{align}
Here we get $a(x)=1,1+u,1+2u ~\text{or}~ 1+3u$. So by the Theorem 3.1, $f_1(x)=f_1^*(x)$ i.e, $f_1(x)$ s self-reciprocal. Also we have, $x^if_2^*(x)=f_2(x)$.
Now suppose, 
\begin{align}
	&
	uf_3(x)=u+uh_1^{\prime}x+\cdots+uh_{k-1}^{\prime}x^{k-1}+ux^k, \nonumber \\
	&
	2uf_4(x)=2u+2uh_1^{\prime\prime}x+\cdots+2uh_{l-1}^{\prime\prime}x^{l-1}+2ux^l, \nonumber \\
\end{align}
where $k>l$. Then
\begin{align}
	uf_3(x) + 2uf_4(x) & = (u+2u) + (uh_1^{\prime}+2uh_1^{\prime\prime})x + \cdots + (uh_{l-1}^{\prime}+2uh_{l-1}^{\prime\prime})x^{l-1} \nonumber \\ & + (uh_l^{\prime}+2u)x^l + uh_{l+1}^{\prime}x^{l+1} + \cdots + uh_{k-1}^{\prime}x^{k-1} + ux^k. \nonumber
\end{align}
\begin{multline}
	\lefteqn (~{uf_3(x)+2uf_4(x)})^{rc} \\ = (1+u)(1+x+\cdots+x^{n-k-2}) + x^{n-k-1} + \overline{uh_{k-1}^{\prime}}x^{n-k} + \cdots + \overline{uh_{l+1}^{\prime}}x^{n-l-2} \\ + \overline{(uh_l^{\prime}+2u)}x^{n-l-1} + \overline{(uh_{l-1}^{\prime}+2uh_{l-1}^{\prime\prime})}x^{n-l} + \cdots +  \overline{(uh_1^{\prime}+2uh_1^{\prime\prime})}x^{n-2} + (1+2u)x^{n-1} \\ = 
	(1+u)(1+x+\cdots+x^{n-k-2}) + x^{n-k-1} + \overline{uh_{k-1}^{\prime}}x^{n-k} + \cdots + \overline{uh_{l+1}^{\prime}}x^{n-l-2} \\ + \overline{uh_l^{\prime}}x^{n-l-1} + 2ux^{n-l-1} + \overline{uh_{l-1}^{\prime}}x^{n-l} + 2uh_{l-1}^{\prime\prime}x^{n-l} + \cdots +  \overline{uh_1^{\prime}}x^{n-2} + 2uh_1^{\prime\prime}x^{n-2} \\ + 2ux^{n-1} + x^{n-1} \\ =
	(1+u)(1+x+\cdots+x^{n-k-2}) + x^{n-k-1} + \overline{uh_{k-1}^{\prime}}x^{n-k} + \cdots + \overline{uh_{l+1}^{\prime}}x^{n-l-2} \\ + \overline{uh_l^{\prime}}x^{n-l-1} + \overline{uh_{l-1}^{\prime}}x^{n-l} + \cdots + \overline{uh_1^{\prime}}x^{n-2} + x^{n-1}  \\ +  3.2ux^{n-l-1} + 3.2uh_{l-1}^{\prime\prime}x^{n-l} + \cdots + 3.2uh_1^{\prime\prime}x^{n-2} + 3.2ux^{n-1} \nonumber
\end{multline}
Since $C$ is linear code, we must have
\begin{equation}
	({uf_3(x)+2uf_4(x)})^{rc} + 3(1+u)\frac{1-x^n}{1-x}\in C.\nonumber
\end{equation}
Hence,
\begin{multline}
	({uf_3(x)+2uf_4(x)})^{rc} + 3(1+u)((1-x^n)/(1-x)) \\ = 
	3ux^{n-k-1} + (\overline{uh_{k-1}^{\prime}}+3(1+u))x^{n-k} + \cdots + (\overline{uh_1^{\prime}}+3(1+u))x^{n-2} + 3ux^{n-1} \\ + 3.2x^{n-l-1}(u+uh_{l-1}^{\prime\prime}x+\cdots+uh_1^{\prime\prime}x^{l-1}+ux^l) \\ =
	3ux^{n-k-1} + 3uh_{k-1}^{\prime}x^{n-k} + \cdots + 3uh_1^{\prime}x^{n-2} + 3ux^{n-1} \\ + 3.x^{n-l-1}(2u+2uh_{l-1}^{\prime\prime}x+\cdots+2uh_1^{\prime\prime}x^{l-1}+2ux^l) \\ =
	3x^{n-k-1}(u+uh_{k-1}^{\prime}x+\cdots+uh_1^{\prime}x^{k-1}+ux^k)\\ + 3.x^{n-l-1}(2u+2uh_{l-1}^{\prime\prime}x+\cdots+2uh_1^{\prime\prime}x^{l-1}+2ux^l) \\ =
	3x^{n-k-1}uf_3^*(x) + 3.2ux^{n-l-1}f_4^*(x)  \\=
	3x^{n-k-1}(uf_3^*(x)+x^{k-l}2uf_4^*(x)) \in C \nonumber.
\end{multline}
Therefore, $f_3(x)=f_3^*(x)$. Also by equation (12) and 
theorem 3.2, we have 
$f_4(x)\vert (2x^jf_{14}^*(x)+2f_{14}(x))$ or $f_4(x)\vert (2x^jf_{14}^*(x)+2f_{14}(x)+2f_2(x))$, \\
where $j=~deg~f_1(x)-deg~f_{14}(x)$. 
Conversely, let $c(x)\in C$, then $c(x)=(f_1(x)+2f_2(x)+2uf_{14}(x))m(x)+(uf_3(x)+2uf_4(x))n(x)$. Since $f_1(x)$ and $f_3(x)$ are self-reciprocal,  $x^if_2^*(x)=f_2(x)$ and $f_4(x)\vert (2x^jf_{14}^*(x)+2f_{14}(x))$ or $f_4(x)\vert (2x^jf_{14}^*(x)+2f_{14}(x)+2f_2(x))$, \\
where $i=~deg~f_1(x)-deg~f_2(x), j=~deg~f_1(x)-deg~f_{14}(x)$, then we write, 
\begin{align}
	c^*(x) & =((f_1(x)+2f_2(x)+2uf_{14}(x))m(x))^*+((uf_3(x)+2uf_4(x))n(x))^* \nonumber \\ &=(f_1^*(x)+2x^if_2^*(x)+2ux^jf_{14}^*(x))m^*(x)+(uf_3^*(x)+2ux^kf_4^*(x))n^*(x) \nonumber \\ 
	& = (f_1(x)+2f_2(x)+2uf_{14}(x))m^*(x)+ (2ux^jf_{14}^*(x)+2uf_{14}(x))m^*(x) \nonumber \\ & + (uf_3(x)+2ux^kf_4^*(x))n^*(x)  \nonumber \\ 
	& = (f_1(x)+2f_2(x)+2uf_{14}(x))m^*(x) + uf_4(x)q(x), ~\text{as}~ f_4(x)\vert f_3(x) \nonumber
\end{align}
Where $c=1,1+u,1+2u ~\text{or}~ 1+3u$. Therefore $c^*(x) \in C$. \\ 
As $3(1+u)((1-x^n)/(1-x)) \in C$, we have
\begin{equation}
	3(1+u)\frac{1-x^n}{1-x}=3(1+u)+3(1+u)x+\cdots+3(1+u)x^{n-1} \in C. \nonumber
\end{equation}
Let $c(x)=c_0+c_1x+\cdots+c_mx^m \in C$. Since $C$ is a cyclic code of length $n$, we have
\begin{equation}
	x^{n-m-1}c(x)=c_0x^{n-m-1}+c_1x^{n-m}+\cdots+c_mx^{n-1} \in C. \nonumber
\end{equation}
Hence, 
\begin{multline}
	\lefteqn ~ 3(1+u)+3(1+u)x+\cdots+3(1+u)x^{n-m-2}+(c_0+3(1+u))x^{n-m-1} \\ +(c_1+3(1+u))x^{n-m}+\cdots+(c_m+3(1+u))x^{n-1} \\ =
	3(1+u)+3(1+u)x+\cdots+3(1+u)x^{n-m-2}+ 3\overline{c_0}x^{n-m-1} \\ +3\overline{c_1}x^{n-m}+\cdots+ 3\overline{c_m}x^{n-1}  \\ =
	3((1+u)+(1+u)x+\cdots+(1+u)x^{n-m-2}+ \overline{c_0}x^{n-m-1} \\ +\overline{c_1}x^{n-m}+\cdots+ \overline{c_m}x^{n-1}) \in C. \nonumber
\end{multline}
This implies $c^*(x)^{rc} \in C$, therefore, $(c^*(x)^{rc})^*=c(x)^{rc} \in C$.

\section{Binary images of DNA codes over $\Z_4+u\Z_4$}
In this Section we will define a Gray map which allows us to translate the properties of the suitable DNA codes for DNA computing to the binary cases. Now we define the Gray map on $R$. Any element $c \in R$ can be expressed as $c=a+ub$, where $a,b \in \Z_4$. The Gray map defined as follows 
\begin{center}
	$\phi:\Z_4 + u \Z_4 \longrightarrow \Z_4^2$\\
	such that $\phi (a+ub)=(b,a+b)$~~~~~~~~$a,b \in Z_4$
\end{center}
Again we give the definition of the Gray map from $\Z_4$ to $\Z_2^2$. First we see that the 2-adic expansion of $c\in \Z_4$ is $c=\alpha(c)+2\beta(c)$ such that $\alpha(c)+\beta(c)+\gamma(c)=0$ for all $c\in \Z_4$. \\
Then we get the table below
\begin{center}
	\begin{tabular}{ l  c  c  c }
		
		$c$&$\alpha(c)$&$\beta(c)$&$\gamma(c)$ \\
		$0$&$0$&$0$&$0$ \\
		$1$&$1$&$0$&$1$ \\
		$2$&$0$&$1$&$1$ \\
		$3$&$1$&$1$&$0$ \\
		
	\end{tabular}
\end{center} 
The Gray map ~~~ $\psi:\Z_4 \longrightarrow \Z_2^2$ ~given by~ $\psi(c)=(\beta(c),\gamma(c))$,~~$c \in \Z_4$ i.e, $\psi(0)=(0,0),\psi(1)=(0,1),\psi(2)=(1,1),\psi(3)=(1,0)$. Now define
\begin{center}
	$\Phi:R\longrightarrow \Z_2^4,~~\Phi=\psi \cdot \phi$\\
	$\Phi(a+ub)=\psi(\phi(a+ub))=\psi(b,a+b)$\\
	$~~~~~~~~~~~~~~=(\beta(b),\gamma(b),\beta(a+b),\gamma(a+b))$.
\end{center}

The Lee weight was defined as $w_L(a)= min \lbrace a,4-a \rbrace, a\in \Z_4$ i.e, $w_L(0)=0,w_L(1)=1,w_L(2)=2,w_L(3)=1$. Let $a+ub$ be any element of the ring $R=\Z_4+u\Z_4, u^2=0$. The Lee Weight $w_L$ of the ring $R$ is defined as follows 
\begin{center}
	$w_L(a+ub)=w_L((b,a+b))$,
\end{center}
where $w_L((b,a+b))$ described the usual Lee weight on $\Z_4^2$. For any $c_1,c_2 \in R$, the Lee distance $d_L$, given by $d_L(c_1,c_2)=w_L(c_1-c_2)$. The Hamming distance $d(c_1, c_2)$ between two codewords $c_1$ and $c_2$ is the Hamming weight of the codeword $c_1-c_2$. It is easy to verify that the image of a linear code over $R$ by $\Phi$ is a binary linear code. In Table 2 we give the
binary image of the codons. In [?] the binary image of DNA code resolved the problem of the construction of DNA codes with some properties. 
\begin{center}
	{\bf Table 2.} Binary Image of the Codons\\~\\
	\begin{tabular}{ l  c  c  c  c  c  c  c  c }
		\hline
		
		$AA$ & $0000$ & $TT$ & $0111$ & $GG$ & $0001$ & $CC$ & $0101$ \\
		$AT$ & $0011$ & $TA$ & $0100$ & $GC$ & $0010$ & $CG$ & $0110$ \\
		$GT$ & $1111$ & $CA$ & $1000$ & $AC$ & $1010$ & $TG$ & $1110$ \\
		$CT$ & $1001$ & $GA$ & $1101$ & $AG$ & $1100$ & $TC$ & $1011$  \\
		\hline
		
	\end{tabular}
\end{center}
The following property of the binary image of the DNA codes comes from the definition. 
\begin{lemma}
	The Gray map $\Phi$ is a distance-preserving map from ($R^n$, Lee distance) to ($\Z_2^{4n}$ , Hamming distance) and this map also $\Z_2$ linear.
\end{lemma}

\begin{lemma}
	If $C$ is a cyclic DNA code of length $n$ over $R$ then $\Phi(C)$ is a binary quasi-cyclic DNA code of the length $4n$ and of index 4.
\end{lemma}
\pf
Let $C$ be a cyclic DNA code of length $n$ over $R$. Hence $\Phi(C)$ is a set of length $4n$ over the alphabet $\Z_2$ which is a quasi-cyclic code of index 4.

\section{Example}
In this section, we give some examples of cyclic codes of different lengths over the ring $R$ to illustrate the above results. 
\begin{example}
	Cyclic codes of length $3$ over $R = \Z_4 + u \Z_4 , u^2 = 0$: We have
	$$ x^3-1 = (x-1)(x^2 + x +1)=g_1g_2 ~\text{over}~ R.$$
	\begin{enumerate}[{\rm (i)}]
	\item Let $C=\langle f_1(x)+2f_2(x)+f_{14}(x) \rangle$, where $f_1(x)=f_2(x)=g_2=(x^2 + x +1)$ and $f_{14}(x)=0$. It is easy to check that $f_1(x)$ is self-reciprocal and $x^if_2^*(x)=f_2(x)$, where $i=~deg~f_1(x)-deg~f_2(x)$. $C$ is a cyclic DNA
	code of length 3 with R-C property and minimum Hamming distance 3. The image of $C$ under the map $\theta$ is a DNA code of length 6, size 16 and minimum Hamming distance 3. These codewords are given in Table 3.
	\item Let $C=\langle f_1(x)+2f_2(x)+2uf_{14}(x),uf_3(x)+2uf_4(x) \rangle$, where $f_1(x)=f_2(x)=g_2, f_3(x)=g_1, f_4(x)=1$ and $f_{14}(x)=0$. We check that $f_1(x)$ and $f_3(x)$ are self-reciprocal, $x^if_2^*(x)=f_2(x)$ and $f_4(x)\vert (2x^jf_{14}^*(x)+2f_{14}(x))$. Then $C$ is a cyclic DNA code of length 3 with R-C property and minimum Hamming distance 3. The image of $C$ under the map $\theta$ is a DNA code of length 6, size 64 and minimum Hamming distance 2.
	\end{enumerate}	 
\end{example}
\begin{center}
	{\bf Table 3.} A DNA code of length 6 obtained from \\ $C=\langle (x^2 + x +1)+2(x^2 + x +1) \rangle$\\~\\
	\begin{tabular}{ l  c  c  c }
		\hline
		
		$AAAAAA$ &  $TTTTTT$ &  $CCCCCC$ &  $GGGGGG$  \\
		$ATATAT$ &  $TATATA$ &  $CTCTCT$ &  $GAGAGA$  \\
		$AGAGAG$ &  $TCTCTC$ &  $CGCGCG$ &  $GCGCGC$   \\
		$ACACAC$ &  $TGTGTG$ &  $CACACA$ &  $GTGTGT$   \\
		\hline
		
	\end{tabular}
\end{center}

\begin{example}
	Cyclic codes of length $7$ over $R = \Z_4 + u \Z_4 , u^2 = 0$: We have
	$$ x^7-1 = (x-1)(x^3 + x +1)(x^3 + x^2 + 1) ~\text{over}~ F_2.$$ 
	This factors are irreducible polynomials over $F_2$. The Hensel lifts of $x^3 + x +1$ to $\Z_4$ is $x^3 + 2x^2 + x - 1$ and Hensel lifts of $x^3 + x^2 +1$ to $\Z_4$ is $x^3 - x^2 - 2x - 1$. Therefore we have
	$$ x^7-1 = (x-1)(x^3 + 2x^2 + x - 1)(x^3 - x^2 - 2x - 1) ~\text{over}~ R.$$
	Let $g_1 = x-1$, $g_2=x^3 + 2x^2 + x - 1$ and $g_3 = x^3 - x^2 - 2x - 1$.\\
	 Let $C=\langle f_1(x)+2f_2(x)+f_{14}(x) \rangle$, where $f_1(x)=f_2(x)=g_2g_3$ and $f_{14}(x)=0$. It is easy to check that $f_1(x)$ is self-reciprocal and $x^if_2^*(x)=f_2(x)$, where $i=~deg~f_1(x)-deg~f_2(x)$. $C$ is a cyclic DNA
	 code of length 7 with R-C property and minimum Hamming distance 7. The image of $C$ under the map $\theta$ is a DNA code of length 14 and minimum Hamming distance 7. The number of codewords are 16 which are given in Table 4.
\end{example}
\begin{center}
	{\bf Table 4.} A DNA code of length 14 obtained from the above code.
	\begin{tabular}{ l  c   }
		\hline
		
		$AAAAAAAAAAAAAA$ &  $TTTTTTTTTTTTTT$ \\ $CCCCCCCCCCCCCC$ &  $GGGGGGGGGGGGGG$  \\
		$ATATATATATATAT$ &  $TATATATATATATA$ \\  $CTCTCTCTCTCTCT$ &  $GAGAGAGAGAGAGA$  \\
		$AGAGAGAGAGAGAG$ &  $TCTCTCTCTCTCTC$ \\ $CGCGCGCGCGCGCG$ &  $GCGCGCGCGCGCGC$   \\
		$ACACACACACACAC$ &  $TGTGTGTGTGTGTG$ \\  $CACACACACACACA$ &  $GTGTGTGTGTGTGT$   \\
		\hline
		
	\end{tabular}
\end{center}

\section{Conclusion} 
In this paper, the algebraic structure of the ring $\Z_4[u]/\langle u^2 \rangle$ and a special family of cyclic codes of odd length over this ring are studied. Cyclic codes is related to DNA codes and their relation is also studied. Reversible codes and reverse-complement codes related to cyclic codes are studied, respectively. Necessary and sufficient conditions for cyclic codes to have the DNA properties have been explored. Again we study binary image of cyclic codes over that ring via the Gray map. For future study, the algebraic structure of cyclic codes of even length and their relation to DNA codes is still an open problem.

%\pagebreak

\end{document}